\documentclass[aps,prl,twocolumn,showpacs,10pt,superscriptaddress,preprintnumbers,nofootinbib]{revtex4-1}
\usepackage{epsfig,amssymb,amsmath,psfrag,epstopdf,color}
\pdfoutput=1
\usepackage{graphicx}
\usepackage{subfigure}
\usepackage[margin=7pt,justification=centerlast]{caption}

\usepackage{mathtools}
\usepackage{xspace}

\usepackage{paralist}
\usepackage{hyperref}
\usepackage{multirow}

\allowdisplaybreaks

\newcommand{\nnlojet}{NNLO\protect\scalebox{0.8}{JET}\xspace}
\DeclareRobustCommand{\LO}{\text{LO}\xspace}

\DeclareRobustCommand{\N}[1]{\ensuremath{\text{N}^{#1}}} 
\DeclareRobustCommand{\ensuremathrm}[1]{\ensuremath{\mathrm{#1}}\xspace}

\DeclareRobustCommand{\ri}{\ensuremathrm{i}} 

\def\beq{\begin{equation}}
\def\eeq{\end{equation}}
\def\bsp#1\esp{\begin{split}#1\end{split}}
\newcommand{\be}{\begin{equation}}
\newcommand{\ee}{\end{equation}}
\newcommand{\bea}{\begin{eqnarray}}
\newcommand{\eea}{\end{eqnarray}}

\def\ksl{\not{\hbox{\kern-2.3pt $k$}}}

\def\spa#1.#2{\left\langle#1\,#2\right\rangle}
\def\spb#1.#2{\left[#1\,#2\right]}
\def\lor#1.#2{\left(#1\,#2\right)}
\def\sand#1.#2.#3{%
\left\langle\smash{#1}{\vphantom1}^{-}\right|{#2}%
\left|\smash{#3}{\vphantom1}^{-}\right\rangle}

\begin{document}

\title{Di-lepton Rapidity Distribution in Drell-Yan Production to Third Order in QCD}
\preprint{KA-TP-17-2021, ZU-TH 33/21, CERN-TH-2021-110, IPPP/21/13}
\author{Xuan~Chen}
\email{xuan.chen@kit.edu}
\affiliation{Physik-Institut, Universit\"at Z\"urich, Winterthurerstrasse 190, CH-8057 Z\"urich, Switzerland}
\affiliation{Institute for Theoretical Physics, Karlsruhe Institute of Technology, 76131 Karlsruhe, Germany}
\affiliation{Institute for Astroparticle Physics, Karlsruhe Institute of Technology, 76344 Eggenstein-Leopoldshafen, Germany}
\author{Thomas Gehrmann}
\email{thomas.gehrmann@uzh.ch}
\affiliation{Physik-Institut, Universit\"at Z\"urich, Winterthurerstrasse 190, CH-8057 Z\"urich, Switzerland}
\author{Nigel Glover}
\email{e.w.n.glover@durham.ac.uk}
\affiliation{
Institute for Particle Physics Phenomenology, Durham University, Durham, DH1 3LE, UK}
\author{Alexander Huss}
\email{alexander.huss@cern.ch}
\affiliation{Theoretical Physics Department, CERN, 1211 Geneva 23, Switzerland}
\author{Tong-Zhi Yang}
\email{toyang@physik.uzh.ch}
\affiliation{Physik-Institut, Universit\"at Z\"urich, Winterthurerstrasse 190, CH-8057 Z\"urich, Switzerland}
\author{Hua~Xing~Zhu}
\email{zhuhx@zju.edu.cn}
\affiliation{Zhejiang Institute of Modern Physics, Department of
  Physics, Zhejiang University, Hangzhou, 310027, China\vspace{0.5ex}}

\begin{abstract}
We compute for the first time the lepton-pair rapidity distribution in the photon-mediated Drell-Yan process to 
 next-to-next-to-next-to-leading order~(\N3\LO) in QCD. The calculation is based on the  $q_T$-subtraction method, suitably extended to this
 order for quark-antiquark initiated Born processes.  Our results display sizeable QCD corrections at \N3\LO  over the full rapidity region and
 provide a fully independent confirmation of the recent results for the total Drell-Yan  cross section at this order. 
 
\end{abstract}

\maketitle


\section{Introduction}
\label{sec:introduction}

Precision physics is becoming increasingly important for the CERN Large Hadron Collider~(LHC) physics program,
in particular in view of the absence of striking signals for beyond the standard model phenomena. Among the most important precision processes at the LHC is Drell-Yan lepton-pair production through neutral current $Z/\gamma^*$ or charged current $W^\pm$ exchanges. It plays a central role in the extraction of standard model parameters and as input to the determination of parton distribution functions (PDFs). The Drell-Yan process is also highly important in new physics searches, both as background to direct signals and as an indirect probe of dynamics beyond the collider energy.   

The Drell-Yan process further plays a special role in the development of modern precision theory calculations in particle physics. It was the first hadron collider process for which next-to-leading order~(NLO) QCD corrections were calculated~\cite{Altarelli:1978id,Altarelli:1979ub}, due to its simplicity in kinematics on one hand and its phenomenological importance on the other hand. The large perturbative corrections observed at the NLO also sparked the interest in soft gluon resummation~\cite{Sterman:1986aj,Catani:1989ne}, which subsequently developed into a field of its own. The first calculation of inclusive next-to-next-to-leading order~(NNLO) QCD corrections to a hadron collider process was also performed for the Drell-Yan process~\cite{Hamberg:1990np,Harlander:2002wh}, followed by the first NNLO rapidity distributions~\cite{Anastasiou:2003yy,Anastasiou:2003ds}, and then fully differential distributions~\cite{Melnikov:2006di,Melnikov:2006kv,Catani:2009sm,Catani:2010en,Gavin:2010az}.  Very recently, next-to-next-to-next-to-leading order~(\N3\LO) QCD corrections have been computed for the inclusive Drell-Yan process with an off-shell photon~\cite{Duhr:2020seh}, and  for charged current Drell-Yan production~\cite{Duhr:2020sdp}. With such level of accuracy in perturbative QCD, mixed electroweak-QCD corrections, derived recently~\cite{Dittmaier:2015rxo,Buccioni:2020cfi,Behring:2020cqi,Buonocore:2021rxx,Bonciani:2021zzf}, become equally important.

For many phenomenological applications of Drell-Yan production, it is more desirable to have differential predictions.
For Higgs production, distributions that are fully differential in the decay products are now available at \N3\LO~\cite{Chen:2021isd} using the analytic results for the inclusive cross section and rapidity distribution of the Higgs boson at \N3\LO as input~\cite{Anastasiou:2015vya,Mistlberger:2018etf,Dulat:2018bfe}. Unfortunately, the same approach does not work well for the Drell-Yan process, as the threshold expansion used for the analytic calculation of the rapidity distribution does not converge well for quark-induced processes.

In this Letter, we present for the first time the di-lepton rapidity distribution for Drell-Yan production at \N3\LO, computed using the $q_T$-subtraction method at this order. We focus on the contribution from virtual photon production alone, neglecting the contribution from $Z$ boson exchange and from virtual photon-$Z$ interference. While the remaining contributions are important, the virtual photon contributions are sufficiently representative to gain knowledge about the size of QCD corrections at this order~\cite{Duhr:2020seh}, and most importantly sufficient for illustrating the subtraction of infrared singularities at \N3\LO.
Upon integration over rapidity, our calculation reproduces the recent \N3\LO result~\cite{Duhr:2020seh} for the
inclusive Drell-Yan coefficient function in a completely independent manner,  
thereby establishing the validity and practicality of the $q_T$-subtraction method at this order.

\section{qT-Subtraction at \texorpdfstring{\N3\LO}{N3LO}}
\label{sec:q_t-subtraction}

The \N3\LO corrections in QCD receive contributions from four types of parton-level sub-processes, each correcting the underlying
Born-level process: 
triple real radiation at tree level, double real radiation at one loop, single real radiation at two loops and purely virtual three-loop corrections. 
At this order, only very few collider processes have been computed so far, 
 including inclusive and differential Higgs production from gluon fusion~\cite{Anastasiou:2015vya,Mistlberger:2018etf,Dulat:2018bfe,Cieri:2018oms,Chen:2021isd,Billis:2021ecs}, inclusive Drell-Yan production~\cite{Duhr:2020seh},
inclusive Higgs production from $b$ quark annihilation~\cite{Duhr:2019kwi}, vector boson fusion Higgs production~\cite{Dreyer:2016oyx}, di-Higgs production~\cite{Chen:2019lzz}, inclusive deep inelastic scattering~\cite{Vermaseren:2005qc} and  
 jet production in deep inelastic scattering~\cite{Currie:2018fgr,Gehrmann:2018odt}.
 
In this Letter, we focus on the Drell-Yan production through a virtual photon, for which all relevant matrix elements have been available for some 
time~\cite{Hagiwara:1988pp,Berends:1988yn,Glover:1996eh,Campbell:1997tv,Bern:1997sc,Garland:2002ak,Baikov:2009bg,Gehrmann:2010ue}. 
After mass factorization of universal initial-state collinear singularities, perturbative predictions for infrared safe observables are finite. 
All individual subprocesses with different multiplicities are separately infrared divergent, with divergences in subprocesses with real radiations
 residing in phase space integrals. An important part of the recent NLO and NNLO revolution has been the development of convenient and efficient algorithms
 for handling these infrared singularities from real emissions. Two among these methods (projection-to-Born~\cite{Cacciari:2015jma} and $q_T$-subtraction~\cite{Catani:2007vq}) have been extended to  be applied in specific \N3\LO
calculations~\cite{Gehrmann:2018odt,Chen:2021isd,Cieri:2018oms,Billis:2021ecs}

The $q_T$-subtraction method~\cite{Catani:2007vq,Catani:2009sm,Catani:2010en,Catani:2011qz} was
  initially developed for processes with colorless final states. 
The key idea is that the most singular phase space configurations are associated with the
small $q_T$ region of the colorless system, and can be isolated by an  artificial $q_T$ cut. The extension of the $q_T$-subtraction
method to \N3\LO has been outlined  for gluon-induced~\cite{Cieri:2018oms,Billis:2019vxg} and quark-induced processes~\cite{Camarda:2021ict,Billis:2019vxg}. 
For Drell-Yan production at \N3\LO, the double differential cross section in di-lepton invariant mass squared $Q^2$ and di-lepton rapidity $y$ is divided into the unresolved~(resolved) part, in which $q_T$ is bounded by $q_T^{\rm cut}$ from above~(below),
\begin{equation}
  \label{eq:2}
  \frac{d^2 \sigma_{\gamma^*}}{dQ^2 dy} =
  \int_0^{q_T^{\rm cut}}\! d^2\boldsymbol{q}_T^{\rm} \frac{d^4 \sigma_{\gamma^*}}{d^2 \boldsymbol{q}_T dQ^2 dy} + 
 \int_{q_T^{\rm cut}} \! d^2\boldsymbol{q}_T^{\rm}  \frac{d^4 \sigma_{\gamma^*}}{d^2 \boldsymbol{q}_T dQ^2 dy} \,.
\end{equation}
The resolved contribution can be regarded as Drell-Yan plus jet production, therefore requiring infrared subtraction only to NNLO. The genuine \N3\LO infrared singularities cancel within the unresolved contribution. While the singularities themselves are canceled, they
give rise to  large logarithms, $\ln^m q_T^{\rm cut}/Q$, both in the resolved and the unresolved contribution, which cancel each other when resolved and unresolved contributions are combined.

A major advantage of $q_T$-subtraction is that the structure of perturbation theory in the unresolved region is well understood from the development of $q_T$ resummation~\cite{Collins:1984kg,Bozzi:2005wk,GarciaEchevarria:2011rb,1202.0814}. This allows one to write the unresolved contributions in a factorised form to all orders in perturbation theory, in terms of a hard function $H$, beam functions $B$ for the incoming particle beams, and a soft function 
$S$:
\begin{align}
  \label{eq:3}
 \frac{d^4 \sigma_{\gamma^*}}{d^2 \boldsymbol{q}_T dQ^2 dy} 
 =&\bigg(
 \sum_i\frac{\sigma_i^{\mathrm{Born}}}{E_{\mathrm{CM}}^2} \int \frac{d^2 \boldsymbol{b}}{(2\pi)^2} \, e^{ i \boldsymbol{q}_T \cdot \boldsymbol{b}}
 \nonumber\\
& \times B_{i/A} (x_A,\boldsymbol{b}) B_{\bar\imath /B} (x_B,  \boldsymbol{b}) S(\boldsymbol{b}) H_i(Q^2) \nonumber\\  &+ (i \leftrightarrow \bar\imath) \bigg)\big[1+ \mathcal{O}({q_T^2/Q^2})\big]\,,
\end{align}
where $\sigma_i^{\mathrm{Born}} = 4 \pi Q_i^2 \alpha_{\mathrm{em}}^2/(3 N_c Q^2)$, $Q_i$ is the electric charge, $\alpha_{\mathrm{em}}$ is the fine structure constant of QED, $E_\mathrm{CM}$ is the center of mass energy. The momentum fractions are
fixed by the final-state kinematics as 
$x_{A} = \sqrt{\tau} e^y$, $x_B = \sqrt{\tau}e^{-y}$, with $\tau = (Q^2 + q_T^2)/E_{\mathrm{CM}}^2$.
In contrast to the leading-power terms~\cite{Ebert:2018gsn,Cieri:2019tfv,Inglis-Whalen:2021bea}, the power corrections are far less well understood but their contribution can be suppressed by choosing a sufficiently small $q_T^{\rm cut}$ value.
The factorisation structure in Eq.~\eqref{eq:3} is most transparent in Soft-Collinear Effective Theory~(SCET)~\cite{hep-ph/0005275,hep-ph/0011336,hep-ph/0109045,hep-ph/0202088,hep-ph/0206152}, which also provides a convenient framework for the calculation of the
unresolved contribution beyond NNLO.   

The hard function $H$ is simply the electromagnetic quark form factor. 
The beam function $B_{i/A}(x_A,\boldsymbol{b})$ encodes initial-state collinear radiation. 
For a high energy hadron $A$ moving in the light-cone direction $n^\mu=(1,0,0,1)$ with four momentum $P_A^\mu$, the beam function can be written in light-cone gauge and coordinates as
\begin{equation}
  \label{eq:6}
  B_{i/A}(x,\boldsymbol{b}) = \int \! \frac{db^-}{4 \pi} e^{i x b^{-}\tfrac{P_A^+}{2}} \langle A|\overline{\psi}_i(0,b^-,\boldsymbol{b}) \frac{\gamma^+}{2} \psi_i(0)  | A \rangle  \,.
\end{equation}
This beam function is a priori a non-perturbative matrix element, which can be expressed in terms of perturbatively calculable Wilson coefficients $I_{i/j}$
and parton distribution functions $f_{j/A}$ using a light-cone operator product expansion:
\begin{equation}
  \label{eq:7}
  B_{i/A}(x,\boldsymbol{b}) = \sum_j \int_{x}^1 \! \frac{d\xi}{\xi} I_{i/j}(\xi,\boldsymbol{b}) f_{j/A}(x/\xi)+ \mathcal{O}(\Lambda_{\text{QCD}}|\boldsymbol{b}|)\,. 
\end{equation} 
The soft function describes multiple soft gluon radiation with a constraint on the total $\boldsymbol{q}_T$. It is given by the vacuum matrix element 
\begin{equation}
  \label{eq:8}
  S(\boldsymbol{b}) = \frac{\mathrm{tr}}{N_c}
\langle \Omega | \mathrm{T}\{Y_{\bar n}^\dagger Y_n(0,0,\boldsymbol{b})\} \overline{\mathrm{T}}\{Y_n^\dagger Y_{\bar n}(0)\} | \Omega \rangle  \,,
\end{equation}
where $Y_n(x) = \mathrm{P} \exp(\ri g \int_{-\infty}^0 ds\, A(x+sn))$ is a path-ordered semi-infinite lightlike Wilson line.

For \N3\LO accuracy, we need the third order corrections to the perturbative beam function $I_{i/j}(x,\boldsymbol{b})$, soft function and hard function. The hard function has been known to three loops for some time~\cite{Baikov:2009bg,1001.2887,Gehrmann:2010ue}. The calculation of the beam and soft function is less straightforward, due to the presence of rapidity divergences~\cite{Collins:2008ht}, which only disappear in physical cross sections. Various approaches for rapidity regularization have been adopted in the literature to obtain the beam and soft function at NNLO~\cite{Catani:2011kr,Catani:2012qa,Gehrmann:2012ze,Gehrmann:2014yya,Echevarria:2016scs,Luo:2019hmp,Luo:2019bmw,Gutierrez-Reyes:2019rug}. At \N3\LO, the scale dependence of perturbative beam and soft functions are completely fixed by renormalisation group~(RG) evolution in SCET; see, e.g.,\ \cite{Chen:2018pzu,Billis:2019vxg}. The initial conditions of this RG evolution form the genuine \N3\LO contributions, and require
calculation to this order in SCET.  Very recently,
this was accomplished in a series of works for the soft function~\cite{Li:2016ctv} and the beam functions~\cite{Luo:2019szz,Ebert:2020yqt,Luo:2020epw}, using the 
rapidity regulator proposed in \cite{Li:2016axz}. These newly available results provide the key ingredients for applying $q_T$-subtraction to processes with colorless final states at \N3\LO. The perturbative beam functions are expressed in terms of harmonic polylogarithms~\cite{Remiddi:1999ew} up to weight 5, which can be 
evaluated numerically with standard tools~\cite{Gehrmann:2001pz}.

The resolved contribution above the $q_T^{\text{cut}}$ for \N3\LO Drell-Yan production contains the same ingredients of the NNLO
calculation with one extra jet. 
Fully differential NNLO contributions for Drell-Yan-plus-jet production have been computed in~\cite{1507.02850,1512.01291,1602.08140}. 
The application to \N3\LO $q_T$-subtraction further requires stable fixed-order predictions at small $q_T$~\cite{1610.01843,1805.05916,1905.05171}, enabling the
cancellation of the $q_T^{\text{cut}}$ between resolved and unresolved contributions to sufficient accuracy.  
In this Letter, we employ the antenna subtraction method~\cite{hep-ph/0501291,hep-ph/0505111,hep-ph/0612257,1301.4693} to compute
 Drell-Yan production above   $q_T^{\text{cut}}$  up to NNLO in perturbation theory, 
 implemented in the parton-level event generator \nnlojet~\cite{1507.02850,1610.01843}. 
To achieve stable and reliable fixed order predictions down to the $q_T\sim 0.4$ GeV region, \nnlojet has been developing dedicated optimizations of its phase space generation based on the work in~\cite{Chen:2018pzu}. This ensures sufficient coverage in
  the multiply unresolved regions required for the $q_T$-subtraction.

\section{Results}
\label{sec:results}
Applying the $q_T$-subtraction method described above, we compute Drell-Yan lepton pair production 
to \N3\LO accuracy. For the phenomenological analysis, we restrict ourselves to the production of a di-lepton pair through a virtual photon only. We take $E_\mathrm{CM} = 13$ TeV as center of mass collision energy  and fix the invariant mass of the di-lepton pair at $Q = 100$ GeV. Central scales for renormalization ($\mu_R$) and factorization ($\mu_F$) are taken at $Q$, allowing us
to compare with the \N3\LO total cross section results from \cite{Duhr:2020seh}. We use the central member of \verb|PDF4LHC15_nnlo| PDFs~\cite{1510.03865}  throughout the calculation.

\begin{figure}[t]
 \centering
\includegraphics[scale=.34]{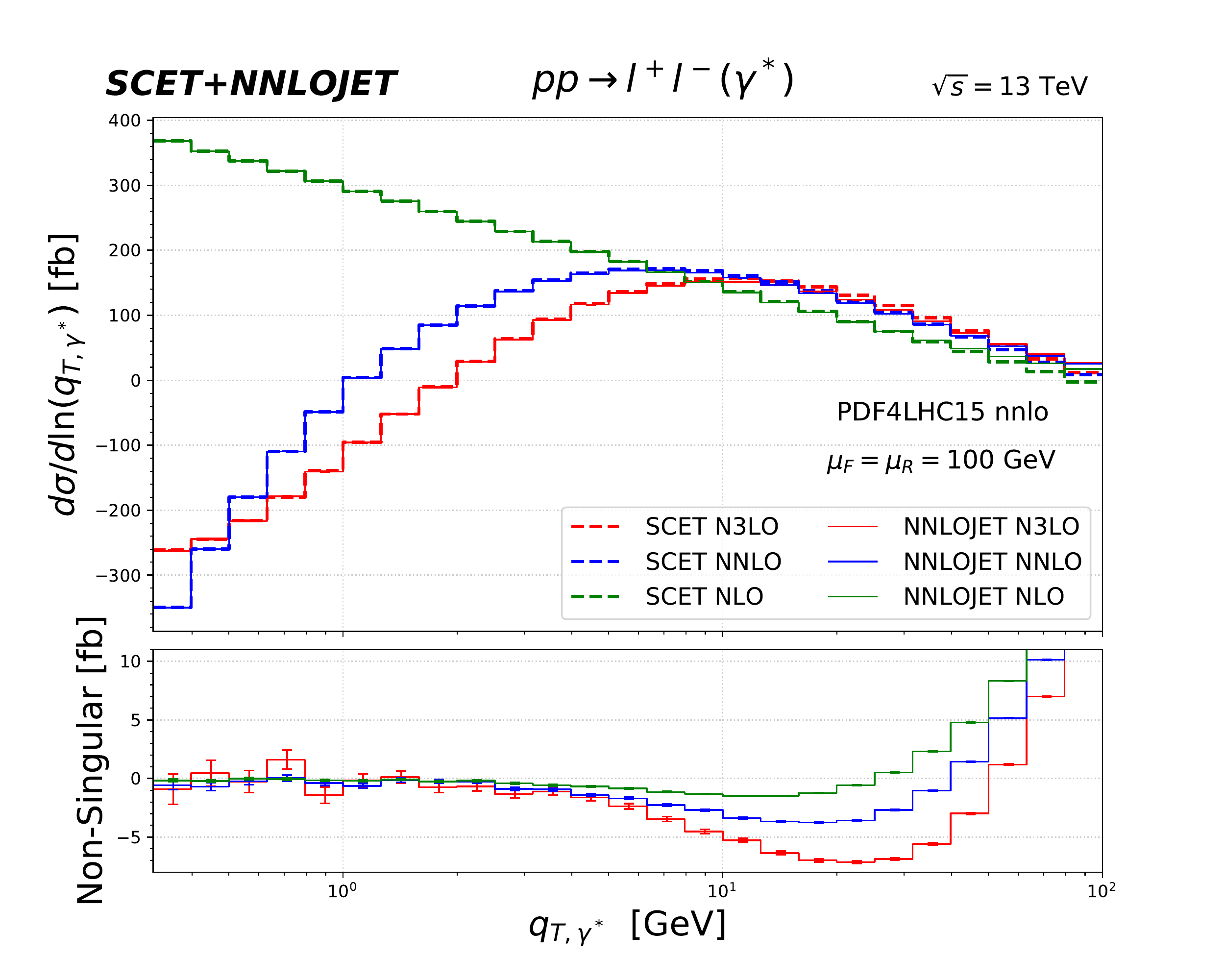}
\caption{\label{fig:gammapT} Perturbative contributions to  transverse momentum distribution of the virtual photon up to $\alpha_s^3$. The upper panel displays the $q_T$-distribution obtained from \nnlojet and from expanding SCET to each order. The bottom panel contains the nonsingular remainder (\nnlojet minus SCET).}
\end{figure}

To establish the cancellation of $q_T^{\rm cut}$-dependent terms between resolved and unresolved contributions, Fig.~\ref{fig:gammapT}
displays the $q_T$ distribution of virtual photon obtained with \nnlojet (used for the resolved contribution) and obtained by expanding
the leading-power factorised prediction at small $q_T$ using Eq.~\eqref{eq:3} up to ${\cal O}(\alpha_s^3)$. 
The highest logarithms at this order are $1/q_T \ln^5(Q/q_T)$. The singular $q_T$ distribution is expected to match between \nnlojet and SCET, which is a prerequisite for the $q_T$-subtraction method. This requirement is fulfilled by the nonsingular contribution (\nnlojet minus SCET) demonstrated in the bottom panel of Fig.~\ref{fig:gammapT}. Remarkably, the agreement starts for $q_T$ at about $2$ GeV and extends down to $0.32$ GeV for each perturbative order. Numerical uncertainties from phase space integrations are displayed as error bars. We emphasize that the observed agreement is highly nontrivial, providing very strong support to the correctness of the \nnlojet and SCET predictions.
 
\begin{figure}[t]
  \centering
  \includegraphics[width=\linewidth]{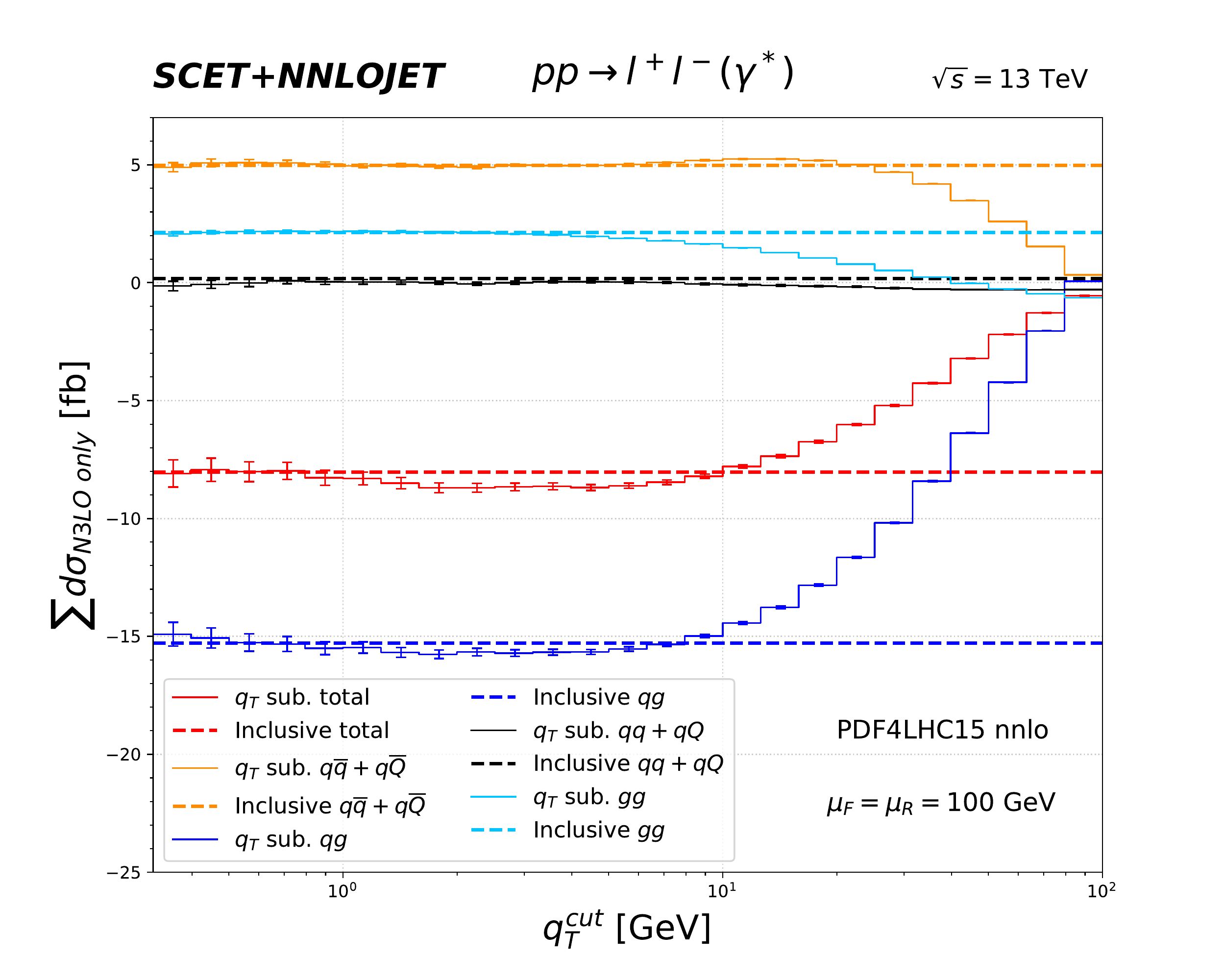}
  \caption{Inclusive \N3\LO QCD corrections to total cross section for Drell-Yan production through a virtual photon.
}
  \label{fig:gammaXS}
\end{figure}

In Fig.~\ref{fig:gammaXS}, we display the \N3\LO QCD corrections to the total cross section for Drell-Yan production through a virtual photon, using the $q_T$-subtraction procedure, decomposed into different partonic channels. The cross section is shown as a function of the unphysical cutoff parameter $q_T^{\rm cut}$, which separates resolved and unresolved contributions.
Integrated over $q_T$, both the \nnlojet and SCET predictions involve logarithms up to $\ln^6 (Q/ q_T^{\rm cut})$, which become explicit in the SCET calculation.
The \nnlojet calculation produces the same large logarithms but with opposite sign, as well as power suppressed logarithms $(q_T^{\rm cut})^m \ln^n(Q/q_T^{\rm cut})$, where $m \ge 2$ and $n \le 6$. The physical
\N3\LO total cross section contribution must not depend on the unphysical cutoff $q_T^{\rm cut}$; therefore it is important to choose a sufficiently small $q_T^{\rm cut}$ to suppress such power corrections.

 Figure~\ref{fig:gammaXS} demonstrates the dependence on $q_T^{\rm cut}$ of the SCET+\nnlojet predictions is negligible for values below $1$ GeV. In fact, for all partonic channels except $qg$, the cross section predictions become flat and therefore reliable already at $q_T^{\rm cut} \sim 5$~GeV. It is only the $qg$ channel that requires a much smaller $q_T^{\rm cut}$, indicating more sizeable power
corrections than in other channels. 

\begin{table}[t]
\centering
  \begin{tabular}{c|c|c|c}
    \hline
    Fixed order & \multicolumn{3}{c}{$\sigma_{pp\rightarrow \gamma^*}$(fb)}\\ \hline
    LO & \multicolumn{2}{c}{\qquad\qquad$339.62^{+34.06}_{-37.48}$}\\ 
    NLO & \multicolumn{2}{c}{\qquad\qquad$391.25^{+10.84}_{-16.62}$}\\ 
    NNLO & \multicolumn{2}{c}{\qquad\qquad$390.09^{+3.06}_{-4.11}$}\\ 
    \N3\LO & \multicolumn{2}{c}{\qquad\qquad\qquad$382.08^{+2.64}_{-3.09}$~\cite{Duhr:2020seh}}\\ \hline 
    \N3\LO only & $q_T^{\text{cut}}=0.63$ GeV & $q_T^{\text{cut}}\rightarrow 0$ fit &~\cite{Duhr:2020seh}\\ \hline
    $qg$  & $-15.32(32)$ & $-15.34(54)$ & $-15.29$ \\ 
    $q\bar{q}+q\bar{Q}$  & $+5.06(12)$ & $+5.05(12)$ & $+4.97$ \\ 
    $gg$  & $+2.17(6)$ & $+2.19(6)$ &$+2.12$ \\ 
    $qq+qQ$  & $+0.09(13)$ & $+0.09(17)$ & $+0.17$ \\ 
    \hline
    Total & $-7.98(36)$ & $-8.01(58)$ & $-8.03$ \\ 
    \hline
  \end{tabular}
   \caption{Inclusive cross sections with up to \N3\LO QCD corrections to Drell-Yan production through a virtual photon. \N3\LO results are from the $q_T$-subtraction method and from the analytic calculation in~\cite{Duhr:2020seh}. Cross sections at central scale of $Q=100$ GeV are presented together with 7-point scale variation. Numerical integration errors from $q_T$-subtraction are indicated in brackets.~\label{tab:totalXS}}
\end{table}

 Also shown in Fig.~\ref{fig:gammaXS} in dashed lines are the inclusive predictions from~\cite{Duhr:2020seh}, decomposed into different partonic channels. We observe an excellent agreement at small-$q_T$ region with a detailed comparison given in Table~\ref{tab:totalXS}. We present total cross sections at small $q_T^{\text{cut}}$ value (0.63 GeV) and results from fitting the next-to-leading power suppressed logarithms with $q_T^{\text{cut}}$ extrapolated to zero.
 This agreement provides a fully independent confirmation of the analytic calculation~\cite{Duhr:2020seh}, and lends 
 strong support to the correctness for our $q_T$-subtraction-based calculation. We observe large cancellations between $qg$ channel~(blue) and $q\bar{q}$ channel~(orange). While the inclusive \N3\LO correction is about $-8$~fb, the $qg$ channel alone can be as large as $-15.3$~fb. Similar cancellations between $qg$ and $q\bar{q}$ channel can already be observed at NLO and NNLO. The numerical smallness of the NNLO corrections (and of
 its associated scale uncertainty) is due to these cancellations, which may potentially lead to an underestimate of theory uncertainties at NNLO.
\begin{figure}[t]
  \centering
  \includegraphics[width=\linewidth]{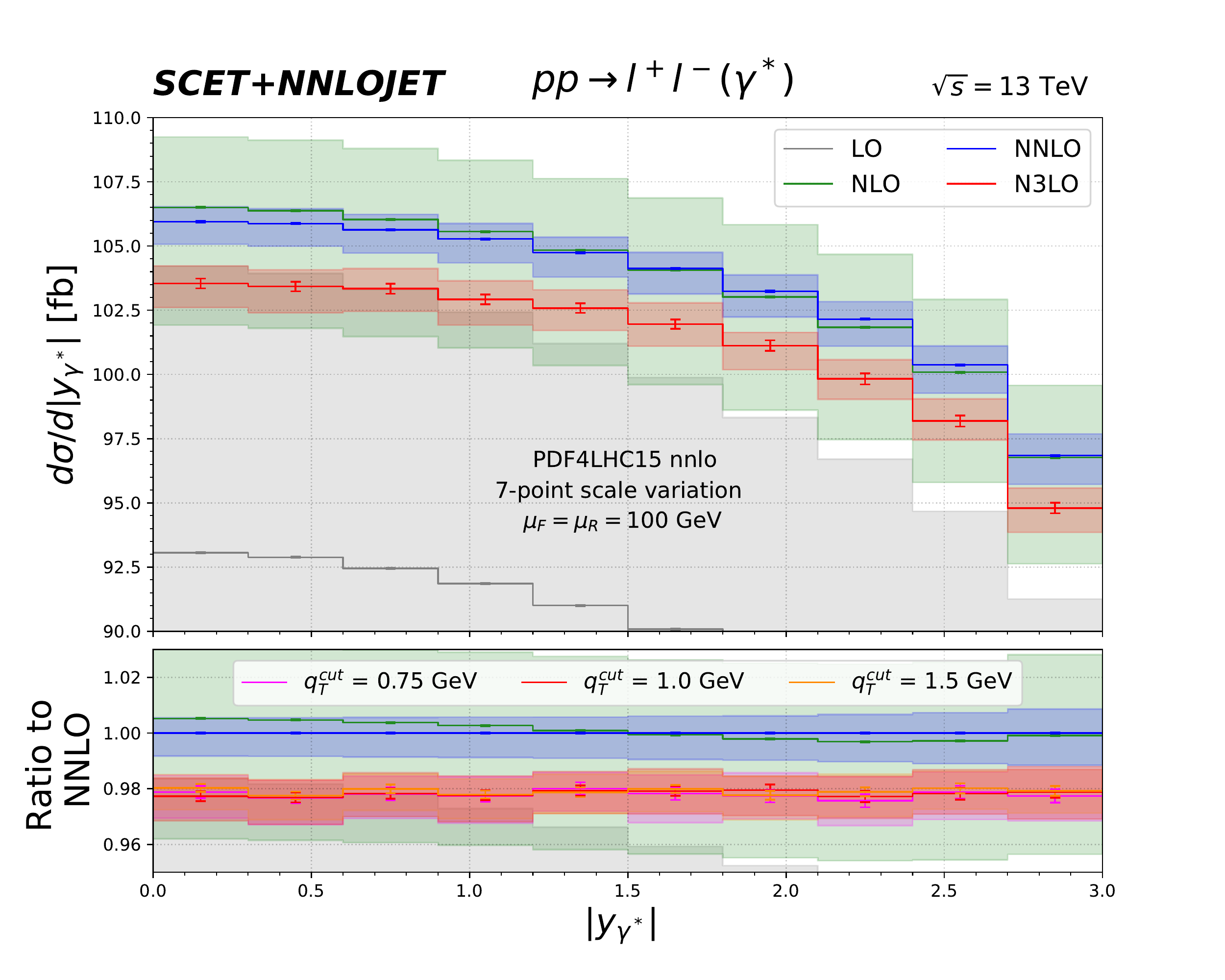}
  \caption{Di-lepton rapidity distribution from LO to \N3\LO. The colored bands represent theory uncertainties from scale variations. The bottom panel is the ratio of  the \N3\LO prediction to NNLO, with different cutoff $q_T^{\rm cut}$.}
  \label{fig:gammaY}
\end{figure}

In Fig.~\ref{fig:gammaY}, we show for the first time the \N3\LO predictions for the Drell-Yan di-lepton rapidity distribution,
which constitutes the main new result of this Letter. Predictions of increasing perturbative orders up to \N3\LO are displayed.
We estimate the theory uncertainty band on our predictions by independently varying $\mu_R$ and $\mu_F$ around 100 GeV with factors of 1/2 and 2 while eliminating the two extreme combinations (7-point scale variation).
With large QCD corrections from LO to NLO, the NNLO corrections are only modest and come with 
scale uncertainties that are significantly reduced~\cite{Hamberg:1990np,Anastasiou:2003yy,Anastasiou:2003ds}. However, as has been observed for the total cross section, the smallness of NNLO corrections is due to cancellations between the $qg$ and $q\bar q$ channels. Indeed, Fig.~\ref{fig:gammaY} shows clearly that the \N3\LO correction is large compared with NNLO,  and that the NNLO scale uncertainty band fails to overlap with \N3\LO over the full rapidity range. It should however be noted that the uncertainties from PDFs, especially from the missing N3LO effects in their evolution, can be at the percent level~\cite{Duhr:2020seh}, which highlights the necessity for a consistent PDF evolution and extraction at \N3\LO in the future.

In the bottom panel of Fig.~\ref{fig:gammaY}, we show the ratio of the \N3\LO rapidity distribution to the previously known NNLO result~\cite{Anastasiou:2003yy,Anastasiou:2003ds}. As can be seen, the corrections are about $-2\%$ of the NNLO results, and are flat over a large rapidity range. There is minimal overlap between the scale uncertainty bands only at large $y_{\gamma^*}$. To test the numerical stability at \N3\LO, three values of $q_T^{\text{cut}}$ are examined in the bottom panel. We observe the $q_T^{\text{cut}}$ dependence to be smaller than the numerical error, which justifies the use of predictions with $q_T^{\text{cut}}=1$~GeV in the top panel.
Since the \N3\LO corrections are largely rapidity independent, their effect will cancel out in the normalized rapidity distribution, which can thus be
expected to be described theoretically to subpercent accuracy, thereby meeting the precision requirements of 
the experimental measurements for normalized distributions in the Drell-Yan process~\cite{1710.05167,1909.04133}.

\section{Conclusion and outlook}
\label{sec:conclusion}

In this Letter, we calculated for the first time the di-lepton rapidity distribution for Drell-Yan production through virtual photon exchange to third order in perturbative QCD. We employed  the $q_T$-subtraction method at \N3\LO, by combining results from NNLO Drell-Yan production at large $q_T$
and leading-power factorised predictions from SCET at small $q_T$. Both contributions are matched at a phase space slicing cut $q_T^{\rm cut}$,
and the cancellation of the leading power $q_T^{\rm cut}$ dependence in the full result provides a strong check. 
Our results firmly establish for the first time the applicability of $q_T$-subtraction at \N3\LO, without any input from a previous inclusive calculation. This opens the door for the application of $q_T$-subtraction at \N3\LO to more complicated final states, either with fiducial final state cuts or for more complex processes.

The newly derived di-lepton rapidity distribution at \N3\LO also opens up an alternative route to
\N3\LO corrections to Drell-Yan type fiducial cross sections. By repeating our calculation for all
Born-type angular coefficients~\cite{Collins:1977iv} in the Drell-Yan process, 
inclusive predictions that are fully differential in the Born-level lepton kinematics can be obtained. These represent the integrated counterterm 
contributions for a fully differential projection-to-Born (P2B) calculation~\cite{Cacciari:2015jma} at \N3\LO~\cite{Currie:2018fgr,Gehrmann:2018odt,Chen:2021isd}.

For total Drell-Yan cross section, our results are in excellent agreement with a previous calculation~\cite{Duhr:2020seh}. We found that \N3\LO corrections are significant over the full rapidity region. They are largely rapidity independent, indicating only very small corrections to distributions that are normalized to the total
cross section. Moreover, perturbative uncertainties estimated from scale variation do not overlap between NNLO and \N3\LO, indicating an underestimate of perturbative uncertainties at NNLO.

To apply our results in  precision phenomenology, one needs to supplement them by contributions from $Z$ boson exchange and $Z$-photon interference. They give rise to new subprocesses taht are infrared finite in the small $q_T$ limit, and therefore one can apply the \N3\LO $q_T$-subtraction method without further modification. It will also be important to combine the QCD results with electroweak corrections and mixed electroweak-QCD corrections. We leave these studies to future work.

\begin{acknowledgments}
\emph{Acknowledgements.}
The authors would like to thank Claude Duhr, Falko Dulat and Bernhard Mistlberger for discussions and Julien Baglio for providing detailed results for the predictions in~\cite{Duhr:2020seh}. This research was supported in part by the Swiss National Science Foundation (SNF) under contract 200020-175595, by the Swiss National Supercomputing Centre (CSCS) under project ID UZH10, and by the Deutsche Forschungsgemeinschaft (DFG, German Research Foundation) under grant 396021762-TRR 257. H. X. Z. was supported by the National Science Foundation of China (NSFC) under contract No. 11975200. N. G. was supported by the U. K. STFC through grant ST/P001246/1.
\end{acknowledgments}

\bibliography{gammaN3LO}

\end{document}